\documentclass[twocolumn, longbibliography
 reprint,
superscriptaddress,
 amsmath,amssymb,
 aps,pra,
]{revtex4-2}

\usepackage{graphicx}
\usepackage{dcolumn}
\usepackage{bm}
\usepackage{braket}
\usepackage{xcolor}
\usepackage{mathtools}
\usepackage{subfigure}
\usepackage{gensymb} 

\usepackage[colorlinks=true, citecolor=blue, urlcolor=magenta]{hyperref}

\begin{document} 


\title{Probing quantum properties of black holes with a Floquet-driven optical lattice simulator} 

\author{Asmae Benhemou}
\thanks{These authors contributed equally.}
\affiliation{Department of Physics and Astronomy, University College London, London WC1E 6BT, United Kingdom}

\author{Georgia Nixon}
\thanks{These authors contributed equally.}
\affiliation{Cavendish Laboratory, University of Cambridge, J.J. Thomson Avenue, Cambridge CB3 0HE, United Kingdom}
\affiliation{School of Physics, University of Sydney, Physics Rd, Camperdown, NSW 2050, Australia}
\author{Aydin Deger}
\affiliation{Department of Physics and Astronomy, University College London, London WC1E 6BT, United Kingdom}
\affiliation{Clarendon Laboratory, University of Oxford, Parks Road, Oxford OX1 3PU, United Kingdom}

\author{Ulrich Schneider}
\affiliation{Cavendish Laboratory, University of Cambridge, J.J. Thomson Avenue, Cambridge CB3 0HE, United Kingdom}

\author{Jiannis K. Pachos}
\affiliation{School of Physics and Astronomy, University of Leeds, Leeds LS2 9JT, United Kingdom} 

\begin{abstract} 
In the curved spacetime of a black hole, quantum physics gives rise to distinctive effects such as Hawking radiation and maximally fast scrambling. 
Here, we present a scheme for an analogue quantum simulation of $(1+1)$ and $(2+1)$-dimensional black holes using ultracold atoms in a locally Floquet-driven optical lattice. We show how the effective dynamics of the driven system can generate position-dependent tunnelling amplitudes that encode the curved geometry of the black hole.
Moreover, we provide a simple and robust scheme to determine the Hawking temperature of a (1+1)D simulated black hole based solely on on-site atom population measurements.  
Combined with the highly tunable onsite atom-atom interactions typical for cold atoms, our simulator provides a powerful and feasible platform to probe the  scrambling of quantum information in black holes. We illustrate the ergodicity of our (2+1)D black-hole simulator by showing numerically that its level statistics in the hard-core limit approaches the ergodic regime faster than a globally homogeneous Hamiltonian. 

\end{abstract}

\maketitle




\section{Introduction} 

A classical particle in the gravitational pull of a black hole cannot escape once it crosses the event horizon. However, quantum fluctuations from the interior of a black hole can tunnel across the horizon and radiate outwards, a phenomenon known as Hawking radiation \cite{Bekenstein1973, hawking_black_1974, hawking1975particle, damour1976black, arzano2005hawking, parikh2000hawking}. The emitted thermal radiation carries no information about the quantum state inside the black hole, and its temperature only depends on the gravitational curvature around the horizon. Note that the Hawking effect is an intrinsic radiation effect (corresponding for example to mass loss for a Schwarzschild black hole), while the related Unruh effect~\cite{unruh1976notes, hu2019quantum}, and its cousin in de Sitter spacetime, the Gibbons-Hawking effect~\cite{gibbons1977action, fedichev2003gibbons}, solely relate to the observer (detector) dependence of the perceived particle content of a given quantum field. The Hawking radiation can be fully analysed in the semiclassical limit ($\hbar$ small), which describes a non-interacting theory of free fields in a static background geometry \cite{Hawking:1980gf, birrell1984quantum}. In addition, any quantum state entering the black hole is expected to scramble (thermalise) with the highest possible rate. This optimal scrambling property \cite{Maldacena-Stanford-BoundChaos-2016} emerges beyond the semiclassical description of black holes, and is the subject of numerous ongoing investigations \cite{Yasuhiro_Sekino_2008, kitaev2015simple, maldacena2016remarks, birrell1984quantum}. 



\begin{figure}[t!]
    \centering
    \includegraphics{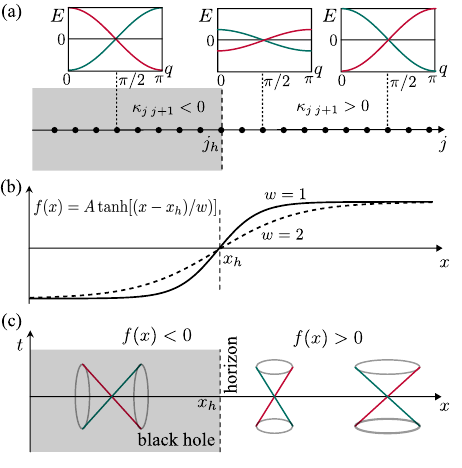}
    \caption{Schematic of the black hole simulator. (a) The black hole geometry is encoded in the position-dependent tunnelling amplitudes $\kappa_{j \> j+1}$ of a fermionic tight-binding chain, and change sign across the horizon. The tunnelling elements $\kappa_{j \> j+1} = [f(jd) +f((j+1)d)]/(8d)$, where $d$ is the discretisation length, follow the behaviour of the black hole curvature $f(x)$, an example of which is given in (b). Insets show the local dispersion relation $E(q)$ at different positions. The left-moving branch is shifted in momentum by $\pi$ to highlight the analogy with light cones. At low energies $|E|\ll J_0$, $E(q)$ is linear and describes massless Dirac fermions in a curved background. (c) The light-cones collapse on each other at the horizon, $x_h=0$, signalling that it is impossible for classical particles to cross. 
    } 
    \label{fig:main_diagram}
\end{figure}

There exist several classical and quantum analogues that can simulate the geometric properties of a black hole, e.g.\ by studying sound waves propagating in a supersonic fluid~\cite{unruh1981experimental, unruh1995sonic, weinfurtner2011measurement, michel2014probing, nguyen2015, coutant2016imprint, bemani2018quantum, jacquet2020next, jacquet2020polariton, jacquet2022analogue, ribeiro2022existence}, using ultracold atoms in the continuum~\cite{lahav2010realization, steinhauer2016observation, munoz2019observation, viermann2022quantum, fischer2004quantum, fedichev2004cosmological}, ion trapping \cite{tian2022testing} or quantum circuits~\cite{jafferis2022traversable}. However, an open question remains as to how to devise a system that can produce the geometrical features of the black hole's horizon, while also exhibiting the scrambling behaviour expected to be present in its interior \cite{sachdev1993gapless, swingle2016measuring, hashizume2021deterministic}.
As an alternative, theoretical proposals have demonstrated how lattice Hamiltonians can simulate quantum field theories in curved space-time geometry~\cite{mallick2019simulating, yang2020simulating, shi_chip_2021, sheng2021simulating,Lapierre2020} and ultracold atoms in optical lattices have emerged as highly flexible and effective analogue quantum simulators of lattice Hamiltonians~\cite{Gross-Bloch-QuantumSimulationsUltracold-2017, Schafer-Takahashi-ToolsQuantumSimulation-2020, Choi-QuantumSimulationsUltracold-2023}. 
Indeed, optical lattices already allow us to probe intriguing quantum many-body phenomena such as entanglement entropy \cite{Islam-Greiner-MeasuringEntanglementEntropy-2015, Kaufman-Greiner-QuantumThermalizationEntanglement-2016},
many-body localisation \cite{ Schreiber-Bloch-ObservationManybodyLocalization-2015, Kondov-DeMarco-DisorderInducedLocalizationStrongly-2015, Choi-Gross-ExploringManybodyLocalization-2016, Bordia-Schneider-CouplingIdenticalOnedimensional-2016, Rispoli-Greiner-QuantumCriticalBehaviour-2019} and quantum scars \cite{Zhao-Knolle-QuantumManyBodyScars-2020, Hudomal-Papic-QuantumScarsBosons-2020}.

Here, we propose to simulate both (1+1)D and (2+1)D black holes using an optical lattice and investigate the resulting behaviour. We employ local Floquet driving to realise position-dependent tunnelling~\cite{Wang-Eckardt-FloquetEngineeringOptical-2018, Nixon-Schneider-IndividuallyTunableTunnelling-2024} in order to encode massless Dirac fermions in the gravitational curvature of a black hole, as shown in Fig. \ref{fig:main_diagram}. 
We show that wave packets of single atoms initialised inside the black hole in 1D eventually tunnel across the horizon and can be witnessed from outside the black hole as thermal Hawking radiation.
Based on the effect local curvature has on the quantum evolution of the system, we propose a straightforward scheme to measure the Hawking temperature from onsite population measurements of atoms. 
Additionally, we demonstrate the resilience of our 1D optical lattice simulator under a wide range of experimental imperfections. 
The simplicity of our scheme paves the way for various extensions. We show how our simulator is generalised to $(2+1)$D black holes by using readily available higher-dimensional optical lattices. Introducing tuneable interactions between the atoms in an optical lattice quantum simulator can be achieved using Feshbach resonances~\cite{Chin-Tiesinga-FeshbachResonancesUltracold-2010}. Numerical simulations of an interacting 2D black hole across a $7 \times 7$ site lattice show level statistics of a diffusive phase. Our simulator provides an opportunity to study the many-body physics of black holes beyond the semiclassical regime, where information scrambling can be witnessed.

\section{Optical lattice simulator of black holes} 

We now present the fermionic lattice Hamiltonian that in the low energy limit is mathematically equivalent to the Dirac equation in the curved space around a black hole \cite{birrell1984quantum, yang2020simulating, shi_chip_2021, Mertens2022, morice2022quantum, horner2022emergent, deger2022ads, horner2023chiral}. Our optical lattice simulator is described by a fermionic tunnelling Hamiltonian given by 
\begin{equation}
    H=-J_0\sum_{\langle i , j \rangle }  \kappa_{ij} (\hat{c}_i^{\dagger}\hat{c}^{}_{j} + \hat{c}_{j}^{\dagger}\hat{c}^{}_i) ,
    \label{Eq:2DTightBinding}
\end{equation}
where the $\hat{c}_j^{\dagger}$($\hat{c}_j$) denote the fermionic creation (annihilation) operators on site $j$, $J_0$ defines a global energy scale, and the $\kappa_{ij}$'s are position-dependent dimensionless tunnelling amplitudes that encode the metric of the curved space. In the 1D model we present, $j \in \{ 1,...,N\}$ where $N$ is the number of sites. When varying the
amplitudes $\kappa_{j \> j+1}$ in a 1D model [Fig.~\ref{fig:main_diagram}(a)] according to a possible metric describing a 1D black hole [Fig.~\ref{fig:main_diagram}(b)], the (local) dispersion relation of the lattice near zero energy mirrors the behaviour of the relativistic light-cones, see Fig.~\ref{fig:main_diagram}(c). In the continuum limit and at half filling, the behaviour of the fermionic chain can be faithfully described by the massless Dirac equation $i\gamma^a e^\mu_a (\partial_\mu +\omega_\mu)\psi = 0$. Here the coordinates are $(x^0,x^1)=(ct,x)$ with $c$ the speed of light, $\gamma^a$ are the two-dimensional Dirac matrices, $e^a_\mu$ are the vierbeins connecting the curved metric $g_{\mu \nu} =\eta_{ab} e^a_\mu e^b_\nu$, to the flat metric $\eta_{ab}=\rm{diag}(1,-1)$, where  $\omega_\mu$ is the spin connection and $\psi$ is the two-component spinor (see Supp. Mat. for details \cite{Supp_Mat} and Refs.~\cite{born1961born, yepez2011einstein} for an introduction to this formalism). The time evolution under the lattice Hamiltonian Eq.~\eqref{Eq:2DTightBinding} is equivalent to that of a Dirac field in a curved background under the metric 
\begin{equation}
ds^2 = f(x) c^2 dt^2 - \frac{1}{f(x)} dx^2,
\label{eq:metric}
\end{equation}
when $\gamma^0=\sigma^z$, $\gamma^1=i\sigma^y$, $\hbar c=J_0$ \cite{pachos2010}, where the spacetime curvature $f(x)$ is encoded in the lattice tunnelling amplitudes via $\kappa_{j \> j+1}= [f(jd)+f((j+1)d)]/(8d)$ and $d$ is a chosen discretisation length \cite{yang2020simulating, Supp_Mat}. A typical Schwarzschild-like metric $f(x)$ in Eq.~\eqref{eq:metric} possesses a horizon at a distance $x_h$ where $f(x_h)=0$, while $f(x) < 0$ describes the black hole interior ($x < x_h$) and $f(x) > 0$ the exterior ($x > x_h$), as shown in Fig.~\ref{fig:main_diagram}(b). In this metric, that corresponds to the free-falling frame, the light cones compress radially near the event horizon due to the strong gravitational pull as depicted in Fig.~\ref{fig:main_diagram}(c). At the horizon, the lightcone collapses to a vertical line, signalling the impossibility for classical particles to cross. Nevertheless, quantum particles can tunnel across the horizon resulting in Hawking radiation. The change of sign of $f(x)$ across the event horizon leads to an apparent interchange of space and time coordinates within the black hole. This is encoded in the corresponding change of sign in the tunnelling amplitudes $\kappa_{j \> j+1}$. 
A typical profile for (1+1)-dimensional black holes is $f(x) = A\tanh(x/w)$ \cite{yang2020simulating, horner2023chiral}, where $w$ is positive and controls the steepness of the curve around the horizon and A is a scaling factor, see Fig.~\ref{fig:main_diagram}(b).  For simplicity, in our proposed simulator we approximate this profile around the horizon with a linear function where the dimensionless slope $\alpha=A d/w$ is controlled by the width $w$. This corresponds to tunnelling coefficients of the form
\begin{equation}
    \kappa_{j \> j+1} = \alpha(j-j_h+0.5)/4,
    \label{Eq:LinearTunnellingProfile} 
\end{equation}
such that $\kappa_{j \> j+1} \neq 0$ at the horizon despite the required $f(x_h) = 0$. 

\begin{figure}[t]
    \centering 
    \hspace{-5mm}\includegraphics{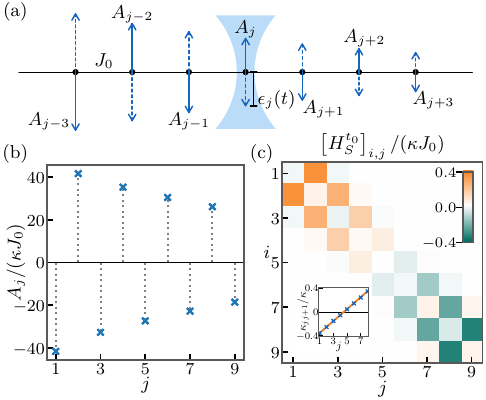}
    \caption{A black hole simulator implemented using local Floquet driving of a 1D optical lattice. (a) The onsite potential of each site $j$ is modulated according to $\epsilon_j(t) = A_j \cos( \omega t) $ with an independent amplitude $A_j$ using individual  tweezer beams indicated by the light-blue shading. 
    (b) Driving amplitudes $A_j$ for a nine-site lattice that generate renormalised dimensionless tunnelling elements $\kappa_{j \> j+1}$ following the linear profile given by Eq.~(\ref{Eq:LinearTunnellingProfile}). The inverted sign of neighbouring driving amplitudes indicates the modulations are $\pi$ out of phase.
    (c) The numerically calculated stroboscopic Hamiltonian $H_S^{t_0}$ for $t_0=0$, $\hbar \omega/\kappa J_0=25$, and the $A_j$ shown in (b). The inset shows that the stroboscopic tunnelling elements perfectly follow the target linear profile in Eq.~(\ref{Eq:LinearTunnellingProfile}) (red line). Here we took $\alpha/\kappa = 0.4$ and $j_h = 5$.} 
    \label{fig:floquet_exposition}
\end{figure}

The discrete lattice representation of a black hole with curvature encoded into the local tunnelling coefficients $\kappa_{j \> j+1}$ in Eq.~\eqref{Eq:2DTightBinding} can be implemented using a locally-driven, one-dimensional optical lattice, where tightly-focused far off-resonant optical tweezers~\cite{  Bernien-Lukin-ProbingManybodyDynamics-2017, Barredo-Browaeys-SyntheticThreedimensionalAtomic-2018, maertens2024hawking, Trisnadi-Chin-DesignConstructionQuantum-2022, Young-Kaufman-Tweezerprogrammable2DQuantum-2022} 
 independently drive the on-site potentials of individual lattice sites in an optical lattice~\cite{Nixon-Schneider-IndividuallyTunableTunnelling-2024}, as illustrated in Fig.~\ref{fig:floquet_exposition}. 
In the following, we utilise Floquet-Bloch theory to capture the long-term evolution of the periodically driven system and demonstrate that it corresponds to a  time-independent effective Hamiltonian \cite{ Goldman-Dalibard-PeriodicallyDrivenQuantum-2014, Bukov-Polkovnikov-UniversalHighFrequencyBehavior-2015, Eckardt-ColloquiumAtomicQuantum-2017, Weitenberg-Simonet-TailoringQuantumGases-2021} that matches the desired position-dependent tunnelling profile given by Eq.~(\ref{Eq:LinearTunnellingProfile}). The full time-dependent Hamiltonian of the driven system is given by
\begin{equation}
    H(t) = -J_0 \kappa \sum_{ \langle i,j \rangle} (\hat{c}_i^{\dagger} \hat{c}_{j} + \hat{c}_{j}^{\dagger} \hat{c}_i ) + \cos (\omega t) \sum_j A_j \hat{c}_j^{\dagger} \hat{c}_{j}.
    \label{Eq:timedependent2dHamiltonian}
\end{equation}
where $\kappa$ is a dimensionless value such that $\kappa J_0$ is the bare tunnelling between neighbouring lattice sites, $A_j$ is the driving amplitude of site $j$, and $\omega$ is the global driving frequency ensuring $H(t) = H(t+T)$ with period $T = 2 \pi /\omega$. The constant $\kappa$ is required to encode a larger range of Hawking temperatures in our simulator \cite{Supp_Mat}. The $T$-periodic nature of Eq.~(\ref{Eq:timedependent2dHamiltonian}) ensures 
that the resulting stroboscopic evolution during multiples of  the global driving period $T$ can be captured by
\begin{equation}
    U(t_0 + nT, t_0) = e^{-\frac{i}{\hbar} nT H_S^{t_0}},
    \label{Eq:StroboscobicTimeEvolution}
\end{equation}
where $n \in \mathbb{N}$, and $H_S^{t_0}$ is the time-independent, ``stroboscopic Hamiltonian" \cite{Eckardt-ColloquiumAtomicQuantum-2017, Bukov-Polkovnikov-UniversalHighFrequencyBehavior-2015}. The stroboscopic Hamiltonian is analytically difficult to calculate, but can be accessed numerically \cite{Supp_Mat}. A convenient analytical approximation of $H_S^{t_0}$ can be obtained in the high-frequency regime, resulting in Hamiltonian Eq.~\eqref{Eq:2DTightBinding} with
\begin{equation}
\kappa_{ij} = \kappa \mathcal{J}_0 \left( \frac{|A_i - A_{j}|}{\hbar \omega} \right),
\label{Eq:KappaFromDrive}
\end{equation}
where $\mathcal{J}_0(y)$ is the zeroth order Bessel function \cite{Supp_Mat}. 
Crucially, Eq.~(\ref{Eq:KappaFromDrive}) illustrates that local tunnelling elements depend only on the relative modulation between neighbouring lattice sites~\cite{Nixon-Schneider-IndividuallyTunableTunnelling-2024}. We can now reverse-engineer Eq.~\eqref{Eq:KappaFromDrive} to generate a sequence of $A_j$ and a global  $\kappa$ that result in the desired tunnelling profile given in Eq.~(\ref{Eq:LinearTunnellingProfile}), and hence encode the black-hole geometry. For example, Figure~\ref{fig:floquet_exposition}(b) shows a series of driving amplitudes $A_j$ that generate $\kappa_{j \>  j+1}$ following the linear profile of Eq.~(\ref{Eq:LinearTunnellingProfile}) with $\alpha/\kappa = 0.4$ and $j_h=5$. In order to minimise the absolute size of the required $A_l$, and thereby the resulting Floquet heating \cite{Weinberg-Simonet-MultiphotonInterbandExcitations-2015, Reitter-Schneider-InteractionDependentHeating-2017, Eckardt-ColloquiumAtomicQuantum-2017}, we alternate their signs. 

Figure \ref{fig:floquet_exposition}(c) shows the numerically generated \cite{Supp_Mat} stroboscopic Hamiltonian obtained from the driving amplitudes of Fig.~\ref{fig:floquet_exposition}(b). This operator matches the analytic approximation in Eq.~(\ref{Eq:KappaFromDrive}) up to small higher-order corrections such as next-to-nearest neighbour tunnelling elements, and new onsite energies that are invisible on this scale~\cite{Nixon-Schneider-IndividuallyTunableTunnelling-2024}. As we demonstrate later, our analogue black hole simulator is resilient to these corrections and the resulting stroboscopic Hamiltonian can faithfully reproduce the expected Hawking radiation.



\subsection{Hawking temperature from radiation population} 

To demonstrate the reliability of our simulator in reproducing the quantum features of a black hole, we first investigate the behaviour of the Hawking temperature
\begin{equation}
     T_H= \frac{\hbar c}{4\pi k_B} \left.\frac{ d f(x)} {dx}\right|_{x=x_h},
\end{equation}
which depends on the curvature of the black hole at the horizon \cite{hawking1975particle}. We then propose a simple method to measure $T_H$ in an optical lattice experiment by utilising easily accessible in-situ density measurements to analyse the quantum walk of an initially localised atom \cite{Dur-Briegel-QuantumWalksOptical-2002, Weitenberg-Kuhr-SinglespinAddressingAtomic-2011, Fukuhara-Gross-MicroscopicObservationMagnon-2013}. In the following, we take for simplicity $\hbar =1$, $k_B =1$ and $c=J_0=1$ unless stated otherwise. This leads to $\alpha = 4\pi T_H$ in our linearised simulator given in \eqref{Eq:LinearTunnellingProfile}.  

\begin{figure}[t!]
    \centering\hspace{-3.5mm}\includegraphics[width=0.47\textwidth]{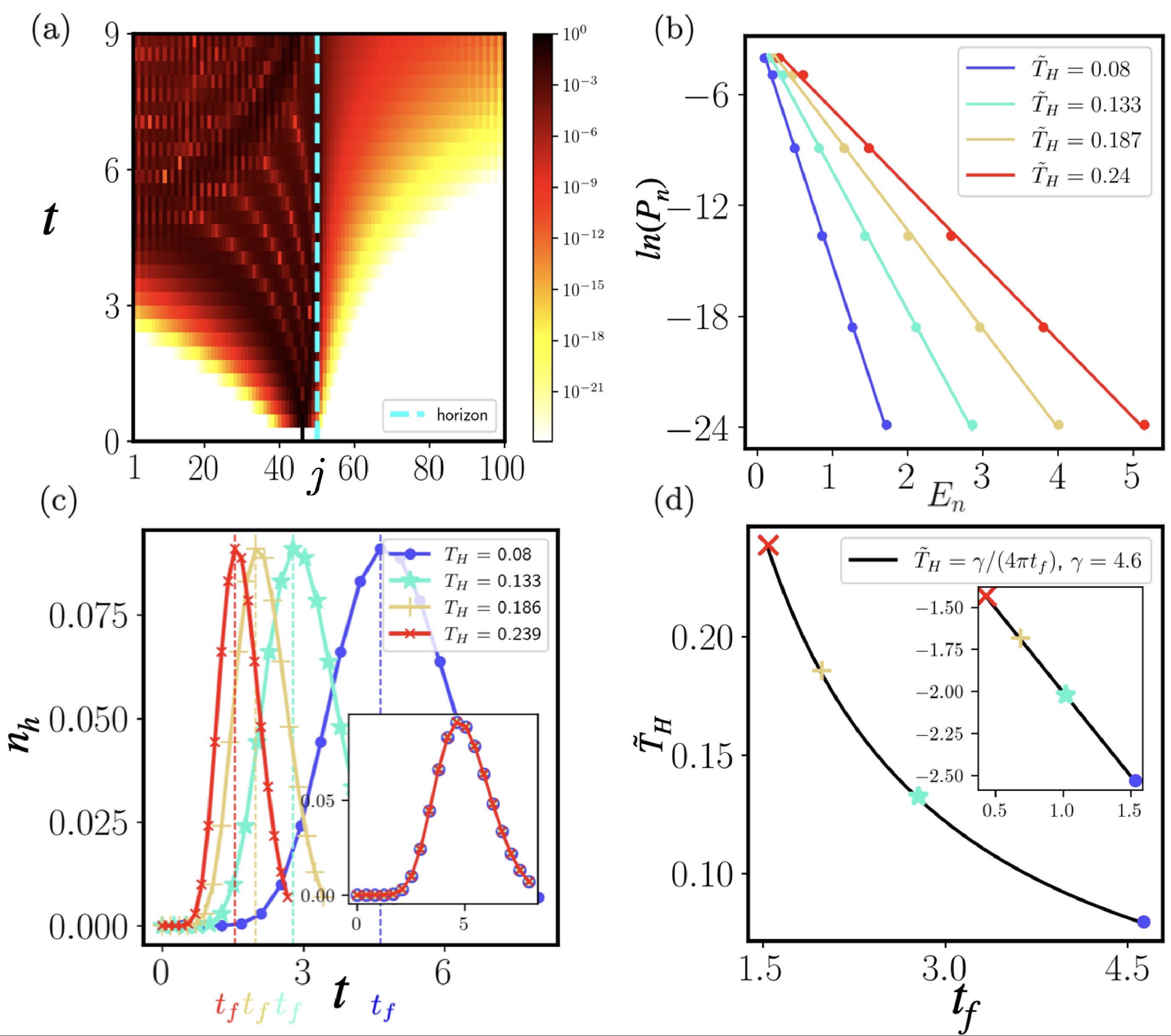}
    \caption{(a) Density evolution (log-scale) of a single particle initialised at $j = 45$ in a chain of length $L = 100$ under the Hamiltonian in Eq.~\eqref{Eq:2DTightBinding} using a  Hawking temperature $T_H = \alpha/(4\pi)\approx0.08$. (b) Probability (log scale) of finding a particle with energy $E_n$ outside the event horizon satisfying a blackbody spectrum $P_n \propto e^{-E_n/\tilde{T}_H}$, evaluated at time $t = 6/\alpha$, for varying $\alpha$. Extracted temperatures $\tilde{T}_H$ from the thermal distribution agree with the expected Hawking temperature $T_H$ encoded in the Hamiltonian. (c) Time evolution of the density $n_{\textrm{h}}$ at the horizon site $j_h=50$ for several $\alpha$ (or $T_H$). Same colors in (b) and (c) corresponds to the same $\alpha$. $t_f$ denotes the time at which peak density is reached at the horizon. (Inset) Collapsed curves with time rescaled by $1/\alpha$. (d) Hawking temperature $\tilde{T}_H$ extracted from spectrum in (b) against the measured time $t_f$ of the peak of the density at the event horizon. The solid line fits the data to the ansatz $\tilde{T}_H \propto t_f^{-1}$ (inset in log-log scale). This relationship allows for the extraction of $\tilde{T}_H$ from an experimentally measured peak time $t_f$.}
    \label{fig:main_population_info_plots}
\end{figure}

To study the physics close to the simulated event horizon we consider the evolution of a single quantum particle, termed the walker, initially positioned inside the black hole, and its potential tunnelling across the horizon.  
In Fig.~\ref{fig:main_population_info_plots}(a) we show the density evolution of such a walker under the linearised tunnelling profile of Eq.~\eqref{Eq:LinearTunnellingProfile} with $j_h = 50$, $\alpha=1$ on a chain of $100$ sites where the walker is initially at site $j=45$, i.e.\ inside the black hole. 
The walker noticeably remains mostly confined inside the black hole but with a non-zero probability of escaping, i.e. \textit{evaporating} \cite{sabsovich2022hawking, horner2022emergent}. This forms the outgoing Hawking radiation.
We explore walker dynamics for a range of Hawking temperatures 
by evolving the initially localised state $\ket{\Psi_i}$ numerically using $\ket{\Psi(t)}=e^{-i\mathcal{H}t}\ket{\Psi_i}$.

An observer outside the horizon only has access to the visible (outside) part of the system and will therefore see the emergence of an incoherent state expected to correspond to a blackbody spectrum at temperature $T_H$ \cite{gibbons1977action}.
We compute the probability $P_n = \bra{E_n} \rho_{\textrm{out}}\ket{E_n}$ of a particle with energy $E_n$ being detected outside the horizon, where $\rho_{\textrm{out}}$ is the density matrix of the evolved state $\rho_t = \ket{\psi(t)}\hspace{-1mm}\bra{\psi(t)}$ at time $t$ truncated at the event horizon, and $\ket{E_n}$ are the energy eigenstates of the Hamiltonian $H_{\rm{out}}$ taken to be acting only on the sites outside the horizon. In analogy with the particle-type solutions of the Dirac equation,  we focus our analysis on positive energies, and indeed find that this spectrum decays exponentially as $P_n \propto e^{-E_n/\tilde{T}_H}$, as shown in Fig.~\ref{fig:main_population_info_plots}(b). 


 In fitting the slopes in Fig.~\ref{fig:main_population_info_plots}(b) we ignored the exponentially suppressed contributions from very large energies as our Dirac simulation is valid in the low energy regime. We also implemented a cut-off of contributions from very small energies where discretisation effects become dominant. While large wavevectors can lead to greybody effects distorting the blackbody spectrum~\cite{holanda2023impact}, the thermal signature remains robust for the energy scale chosen in Fig.~\ref{fig:main_population_info_plots}(b). 
 
Experimentally, directly monitoring the populations of all eigenstates $\ket{E_n}$ necessitates full state tomography in the black hole exterior, which is a tantalising task. 
Using the full tanh profile sketched in Fig.~\ref{fig:main_diagram}(b), it would also be possible to extract the spectrum from the experimentally accessible momentum distribution of the walker in the flat region of space. However, this would require large systems that might be challenging to realise experimentally.
Hence, in the following we propose an alternative means to extract the Hawking temperature, and hence the curvature at the horizon,  purely from in-situ population measurements routinely available in quantum gas microscopes \cite{Bakr-Greiner-QuantumGasMicroscope-2009, Sherson-Kuhr-SingleatomresolvedFluorescenceImaging-2010, Preiss-Greiner-StronglyCorrelatedQuantum-2015, Gross-Bakr-QuantumGasMicroscopy-2021}.
Intuitively, due to the decreasing tunnelling amplitudes close to the horizon, the walker will slow down in this region, before either being reflected back into the black hole or tunnelling through to the exterior, resulting in a pronounced maximum in density that depends on $\alpha$, i.e.\ on the local curvature at the horizon.

To demonstrate this, we plot in Fig.~\ref{fig:main_population_info_plots}(c) the time evolution of the population density $n_h=\langle c_{j_h}^\dagger c_{j_h}\rangle$ at the event horizon site $j_h$.
The density $n_h$ is initially zero until the wave packet has reached the horizon, after which it reaches a maximum at a time $t_f$ before decreasing again. 
In Fig.~\ref{fig:main_population_info_plots}(d) we plot $t_f$ against the temperature $\tilde{T}_H$ extracted from the thermalised spectrum and find the simple inverse relationship 
\begin{equation}
    t_f = \gamma (4\pi \tilde{T}_H)^{-1}
    \label{eq:pop_relationship}
\end{equation}
with a proportionality constant $\gamma$, directly linking Hawking temperature and time dynamics. This effect also transpires in the inset of Fig.~\ref{fig:main_population_info_plots}(c), where the density evolutions at the event horizon are collapsed by a linear rescaling of the time axis. Note that $\gamma$ depends on the particular experimental conditions, in particular the initial position of the walker. 
We highlight that monitoring density at a single site is sufficient since the Hawking temperature is controlled by the local curvature at the horizon~\cite{hawking1975particle, damour1976black, parikh2000hawking, arzano2005hawking, liberati2020back}. Alternatively, one can measure the total probability density outside the horizon, i.e. $P_{\textrm{out}} = \sum_{j=j_h+1}^{j=L} n_j$, where $n_j=\langle c_j^\dagger c_j\rangle$, which provides a direct measure for the total intensity of the Hawking radiation. 
Within the linear profile of Eq.~\eqref{Eq:LinearTunnellingProfile}, scaling the encoded Hawking temperature $T_H$  amounts to an equivalent global scaling of the tunnelling coefficients $\kappa_{j \> j+1}$, which is directly equivalently to an inverse scaling of time and hence explains the behaviour observed in \eqref{eq:pop_relationship}.

So far we have only considered the single-particle case and would hence not yet expect any thermalisation or scrambling. However, due to the specific tunnelling profile that simulates the black-hole horizon, those components of the quantum state that tunnel through the horizon do appear thermal for the degrees of freedom outside the black hole even for a pure initial quantum state. This corresponds to the common notion that the Hawking radiation loses memory of the initial state of the black hole when observed from the outside. 
This is manifested in our simulator by obtaining a thermal spectrum outside the black hole with the same Hawking temperature regardless of the initial position of the walker inside the black hole. 

\begin{figure}[t!]
    \centering
    \hspace{-4mm}\includegraphics[width=\linewidth]{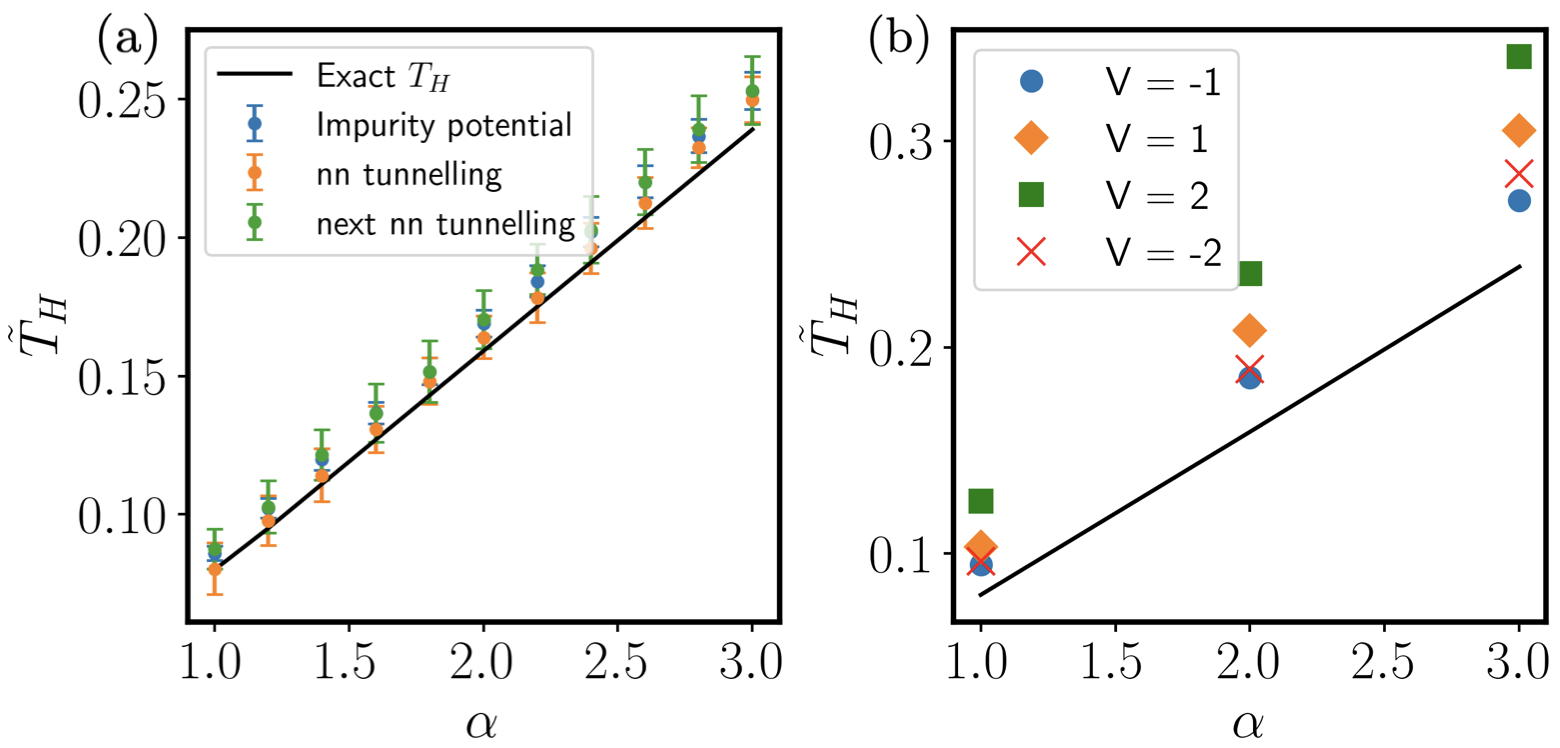}
    \caption{(a) Error resilience to random on-site potentials (blue) with $\varepsilon=0.1$, additive random nearest-neighbour (nn) tunnelling error (orange) ($\varepsilon=0.1$)  and  random next-nearest neighbour (nnn) tunnelling with $\varepsilon=0.01$ (green). These results are averaged over time, initialised sites and 100 realisations for a system with size $L=100$ and horizon $j_h=50$. The error bar indicates the standard deviation across different realisations. The theoretical Hawking temperature value is represented by the black lines. (b) Density-density interaction of strength $V$ for a system size of $L=30$ and $j_h=15$. 
    }
    \label{fig:error}
\end{figure}


\section{Resilience to errors} 

We now investigate the sensitivity of the extracted Hawking temperature to errors pertinent to Floquet-driven optical lattice experiments. Specifically, we extract the Hawking temperature $\tilde{T}_H$ as per the method used in Fig.~\ref{fig:main_population_info_plots}(b), after introducing noise to the system Hamiltonian in Eq.~\eqref{Eq:2DTightBinding}. We model random on-site potentials as an additional term of the form $J_0\sum_{j=1}^N \varepsilon_j \hat{n}_j$ in the Hamiltonian with the coefficients $\varepsilon_j$ drawn randomly from a standard normal distribution with mean zero and variance $\varepsilon$. Despite the existence of random potentials, the essential features of the extracted $T_H$ persist even with an impurity strength of $\varepsilon=0.1$ for a wide range of $\alpha$'s, as shown in Fig.~\ref{fig:error}(a). We next evaluate nearest-neighbour ($k=1$), and next-nearest neighbour ($k=2$) tunnelling errors, represented by $J_0\sum_{j=1}^N \varepsilon_j (\hat{c}_j^{\dagger}\hat{c}_{j+k} + \hat{c}_{j+k}^{\dagger}\hat{c}_j)$. These errors can reflect higher-order corrections beyond Eq.~\eqref{Eq:KappaFromDrive}, see~\cite{Supp_Mat}. Fig.~\ref{fig:error}(a) shows that the extracted Hawking temperature remains stable for error strength $\varepsilon=0.1$ for $k=1$ and $\varepsilon=0.01$ for $k=2$. The observed robustness against all these errors stems from the independence of Hawking temperature on various details of the black hole. 

Finally, we extend our analysis to include two-particle dynamics. Fig.~\ref{fig:error}(b) depicts the outcome of introducing an interaction term $J_0 V \sum_{j=1}^N \hat{n}_j \hat{n}_{j+1}$ between two particles. Our results underscore the resilience of our method against attractive interactions, i.e.\ negative values of $V$, which encourage particle clustering, causing them to act as a single entity, effectively replicating the single-particle behaviour. In contrast, for repulsive interactions ($V=1,2$), the Hawking temperature consistently surpasses its theoretical values across various ranges of $\alpha$.

\section{2D Implementation of the black hole simulator}

The Floquet simulator can directly be extended to a two-dimensional configuration. This allows probing of non-trivial quantum features that are not present in the one-dimensional case. This includes hallmarks such as the area law for the Bekenstein-Hawking entropy, namely that the black-hole entropy scales with the size of its boundary according to thermodynamic arguments \cite{bombelli1986, srednicki1993, CARDY1986186, Andrew_Strominger_1998, Roberto_Emparan_2006}. This property also arises in the Ryu–Takayanagi formula conjectured to describe the relationship between the entanglement entropy of a conformal field theory and the geometry of an Anti-De-Sitter spacetime, i.e.\ the AdS-CFT correspondence \cite{ryu2006holographic, ryu2006aspects, rangamani2017holographic, maldacena1999large, chen2022quantum}. In combination with the contact interactions readily available with cold atoms, the 2D situation also enables studying the interplay between the locally chaotic, thermalising dynamics of a 2D Hubbard model~\cite{rigol2008thermalization,Schneider2012}  with the unique global features of the simulated black-hole horizon.

\begin{figure*}[ht!]
    \centering
\hspace{-5mm}
\includegraphics{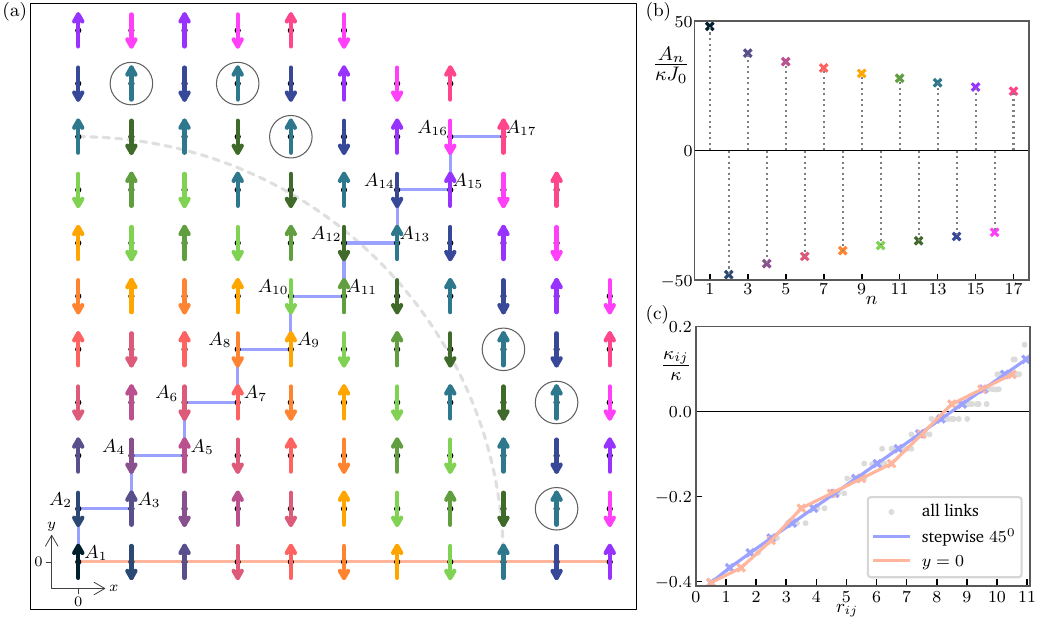}
    \caption{ (a) Analogue quantum simulation of a 2D black hole using a Floquet-driven square optical lattice. Sites are driven with amplitudes $A_n$ where $n \in \{1, ..., 17\}$ as labelled and arrows of the same colour indicate the same amplitude. Neighbouring sites are driven $\pi$ out of phase as depicted by opposite arrow directions.  The simulated event horizon is at the dashed grey line and the centre of the black hole $(x,y) = 0$ is at the site driven by the black arrow with amplitude $A_1$.   Pictured is the top right quarter of the simulator; the simulation is symmetric under $90\degree$ rotations around the centre.  (b) The driving amplitudes indexed by $n$ are chosen such that the tunnelling amplitudes along the $45\degree$ path (blue line in (a)) follow a linear gradient. The alternating sign of $A_n$ shown in (b) indicates that neighbouring sites are driven $\pi$ out of phase. (c) The resulting tunnelling coefficients $\kappa_{ij}$ according to Eq.~\eqref{Eq:KappaFromDrive} vs. the Euclidean distance $r_{ij}$ of each tunnelling link to the black hole centre. The tunnelling coefficients for the blue line give a perfectly linear profile. Tunnelling coefficients following the  $x$ axis (orange line) are also approximately linear. 
    Sites marked by a black circle in (a) have been changed to ensure that all plaquettes have an even number of negative tunnelling links and hence zero flux.
    }
    \label{Fig:2DFloquet}
\end{figure*}

We can simulate a 2D black hole at the origin by approximately encoding a linear curvature profile  $f(r)=\alpha (r-r_h)$, where $r$ denotes the Euclidean distance from the centre of the black hole and the event horizon is at $r_h$.  The combined index $j = (x_j, y_j)$ now corresponds to both $x$ and $y$ coordinates of site $j$. The resulting effective Hamiltonian is then again given by Eq.~\eqref{Eq:2DTightBinding}, but with tunnelling along two directions. 
The renormalised tunnelling coefficients then follow $\kappa_{ij}\propto (r_{ij}-r_h)$, where   $r_{ij} = \sqrt{(x_i/2 + x_j/2) ^2 + (y_i/2 + y_j/2)^2}$ is the distance between the tunnelling link connecting sites $i$ and $j$ and the black-hole centre.

As an example, Fig.~\ref{Fig:2DFloquet} illustrates an implementation of a 2D black-hole simulator on a square lattice using a Floquet-driven 2D lattice as given by Eq.~\eqref{Eq:timedependent2dHamiltonian} with appropriately chosen driving amplitudes $A_j$. To estimate the required amplitudes, we first fix the tunnelling elements $\kappa_{ij}$ on the blue path along the diagonal in Fig.~\ref{Fig:2DFloquet} to match the desired linear gradient $\kappa_{ij}=\alpha (r_{ij}-r_h)$. For the $N$ sites along the diagonal, Eq.~\eqref{Eq:KappaFromDrive} provides a set $\{A_n\},n\in\{1,\dots,N\}$ of $N$ corresponding driving amplitudes.  Similarly to 1D, neighbouring sites are driven out of phase to each other to minimise the required driving amplitudes, see  Figure~\ref{Fig:2DFloquet}(b).
For all remaining sites, we pick that amplitude from the above set $\{A_n\}$ that corresponds to the site on the diagonal blue path with the closest matching Euclidean distance to the centre.
In order to avoid creating unwanted artificial gauge fields~\cite{Nixon-Schneider-IndividuallyTunableTunnelling-2024}, the number of links with negative tunnelling amplitudes on each square plaquette is required to be even, i.e.\ 0, 2, or 4.
As a last step, certain sites just outside of the event horizon (marked by black circles in Figure~\ref{Fig:2DFloquet}(a)) are hence changed to ensure this condition.
Figure~\ref{Fig:2DFloquet}(c) shows the resulting tunnelling amplitudes for all links as a function of the Euclidean distance from the link to the centre of the black hole, highlighting the approximate linear relationship.



The Hawking temperature $T_H$ is encoded in the 2D tunnelling profile analogously to the 1D case, i.e.\ by the prefactor $\alpha = 4\pi T_H$. In a square lattice, we do not expect to perfectly preserve the  rotational symmetry of the black hole. 
However, in the limit of large $r_h$ or small $d$, the small distortions visible in Figure~\ref{Fig:2DFloquet}(c) will disappear. An analytical and numerical investigation of similarly encoded linear tunnelling profiles and the robust validity of the resulting encoded Hawking temperature is given in  Ref.~\cite{deger2022ads}.

\section{Chaotic behaviour}

A central but conjectured property of real black holes is that they are the fastest scrambling---and hence most chaotic---systems in nature \cite{Yasuhiro_Sekino_2008,Shenker-Stanford-BlackHolesButterfly-2014,Maldacena-Stanford-BoundChaos-2016}, a property stemming from the interactions between the particles in the black hole. Crucially, our simulation platform can be extended to interacting quantum particles in the curved spacetime of a black hole by making use of the widely tunable contact interactions natural for fermionic and bosonic cold atoms~\cite{Chin-Tiesinga-FeshbachResonancesUltracold-2010}. This leads to a Hubbard model with an on-site interaction term in addition to the spatially-dependent hopping term from Eq.~\ref{Eq:2DTightBinding}~\cite{bloch2008many}. Interacting 2D Hubbard models are non-integrable and generally show thermalising, ergodic dynamics~\cite{rigol2008thermalization, Schneider2012, Ronzheimer2013, Karamlou2024}.

While the experimental implementation can directly use fermionic or bosonic atoms with tunable interactions,  we now consider hard-core bosons, i.e.\ bosonic atoms in the strongly-interacting limit where the interactions suppress site occupations higher than one, for numerical convenience. In this limit, the Bose-Hubbard Hamiltonian is reduced back to the form of Eq.~\ref{Eq:2DTightBinding}, but with the ladder operators $\hat{c}_i^{\dagger}$, $\hat{c}_j$ now describing hard-core bosons with the commutation relations
\begin{align}
    &[c_i, c_j^\dagger] = [c_i, c_j] = [c_i^\dagger, c_j^\dagger] = 0, \quad \forall i \neq j \\
    &\{c_i, c_i^\dagger\} = 1, \quad 
    (c_i)^2 = (c_i^\dagger)^2 \quad \forall i.
\end{align}
This model is expected to be ergodic in the homogeneous case where all tunnellings are equal, corresponding to flat space~\cite{rigol2008thermalization, Ronzheimer2013, Karamlou2024}. 
To demonstrate the ergodicity of this model also in the black-hole setting, we turn to its level statistics and analyse the ratio between energetically adjacent energy gaps as introduced in Ref.~\cite{oganesyan2007localization}. Namely we calculate
    \begin{equation}
        r_n = \frac{\text{min}(g_n,g_{n-1})}{\text{max}(g_n,g_{n-1})},
        \label{eq:stats-mesure}
    \end{equation}
where $g_n = E_{n+1} - E_n \geq 0$ is the gap between two adjacent eigenvalues of the Hamiltonian at a given fixed number of particles, giving $r_n \in [0,1]$.
In an ergodic system in the thermodynamic limit, the distribution of $r_n$ will show level repulsion and follow the Wigner-Dyson statistics characteristic of the Gaussian Orthogonal Ensemble (GOE) of random matrix theory \cite{Berry-Tabor-LevelClusteringRegular-1977,Bohigas-Schmit-CharacterizationChaoticQuantum-1984,Guhr-Weidenmuller-RandommatrixTheoriesQuantum-1998,oganesyan2007localization}. However, in an integrable system where one would not expect scrambling,  there is no level repulsion and the resulting distribution is Poissonian.

\begin{figure}
    \centering
    \includegraphics[width=1.0\linewidth]{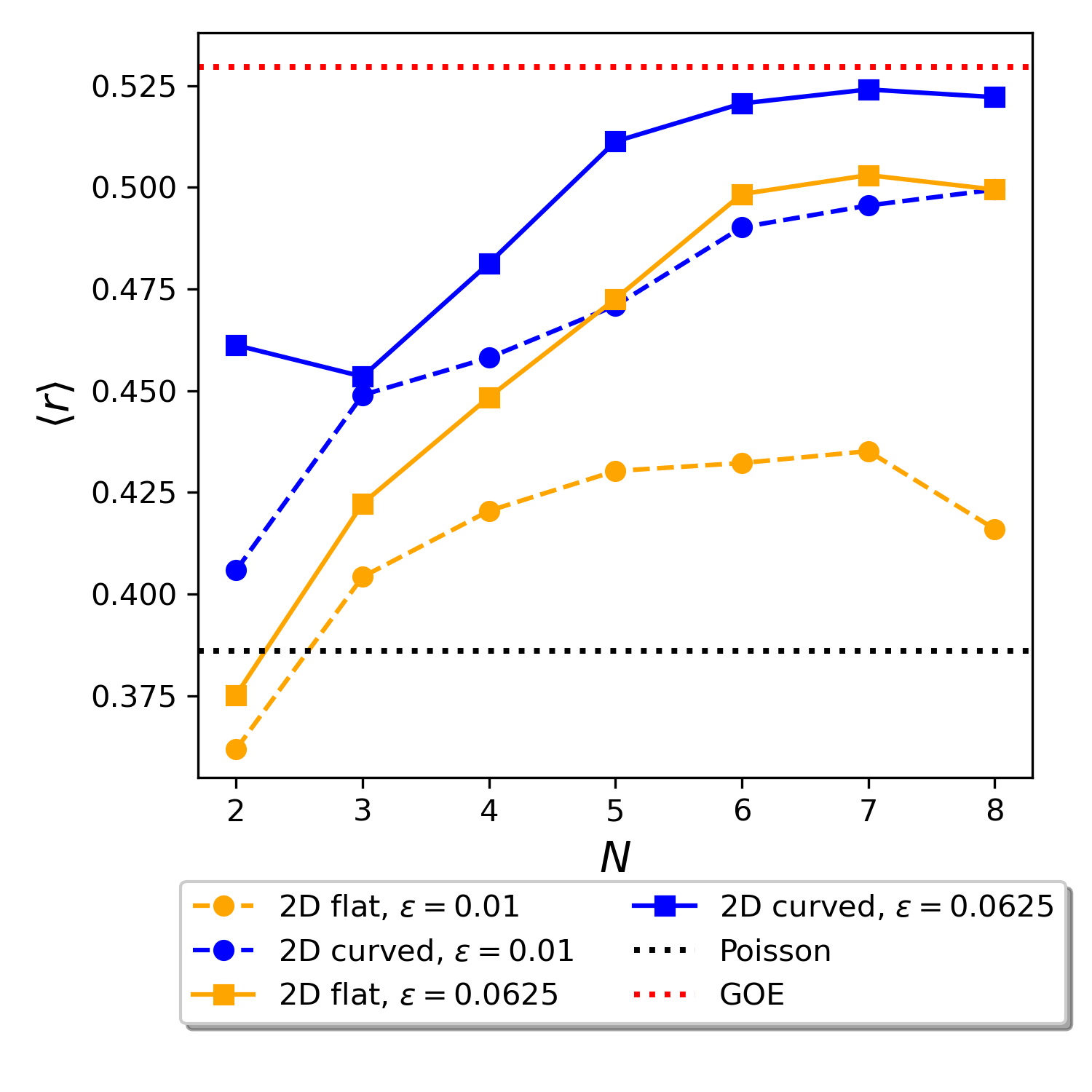}
    \caption{Comparing level statistics between finite 2D square lattices encoding flat (orange) and curved (blue) space-time backgrounds. The mean gap ratio $\langle r \rangle$ is computed using Eq.~\ref{eq:stats-mesure} for varying particle number $N$. For the blue data, $\alpha = 0.46$, $J_0=1$ and we add a perturbation of strength $\epsilon=0.01$ (dashed lines) or $\epsilon=0.0625$ (solid lines) to lift degeneracies. For the flat system, homogeneous tunnelling is set to $\kappa J_0 = 0.28$. The expected mean gap ratios $\langle r_{\text{GOE}} \rangle = 0.5295 \pm 0.0006$ and $\langle r_{\text{Poiss}} \rangle \approx 0.386$ \cite{oganesyan2007localization} for a GOE and Poissonian ensemble in the thermodynamic limit are indicated by red-dotted and black-dotted lines. The curved spacetime encoded in the blue data gives rise to higher values of $\langle r \rangle$ and hence does not break the ergodicity expected for the flat system.}
    \label{fig:fig6}
\end{figure}

Figure~\ref{fig:fig6} compares the mean adjacent energy gap ratios $\langle r \rangle = \frac{1}{s-2} \sum_{n=1}^{s-1} r_n$, where $s$ is the Hilbert space dimension, between multiple 2D systems for varying particle number $N$. The orange data corresponds to a system with homogeneous tunnelling, i.e.\ encoding a flat spacetime, while blue data shows the black hole case with site-dependent tunnellings following Eq.~\eqref{Eq:2DTightBinding}. Both the curved and flat systems effectively simulate an $8 \times 8$ lattice using cone-like boundary conditions, see \cite{Supp_Mat}. In a finite-sized system, Hamiltonian symmetries cause degeneracies in the spectrum, perturbing the level statistics. To circumvent this, we introduce a local potential of size $\varepsilon$ on a single site of the lattice \cite{Supp_Mat}.

The flat system is expected to be ergodic in the thermodynamic limit~\cite{Ronzheimer2013, Karamlou2024}. The data in Figure~\ref{fig:fig6} shows that our black-hole simulator gives level statistics that are consistently closer to an ergodic distribution  compared to the flat system for the same parameter values, thereby demonstrating the potential to study chaotic and scrambling dynamics in this platform.


\section{Conclusions} 

Our study utilises the power of locally driving the sites of an optical lattice to generate an efficient analogue quantum simulator for the intricate physics of Dirac particles in a black hole in one and two spatial dimensions. We introduced a direct approach for determining the Hawking temperature of the simulated black hole from readily accessible onsite atom population measurements at the horizon. The simulated Hawking temperatures are not directly related to the physical temperature of the optical lattice, but can be varied across a broad range by tuning system parameters. This versatility enables the exploration of various scrambling and thermalisation regimes, independent of the limitations imposed by experimental temperature conditions. 
The resulting Hawking temperatures exhibit a remarkable resilience against the common sources of errors inherent in Floquet-driven optical lattices, including the presence of temperature effects, as corroborated by our numerical findings. 

 Furthermore,  optical lattices offer a natural platform for implementing tuneable (non-gravitational) interactions \cite{Chin-Tiesinga-FeshbachResonancesUltracold-2010}. Our simulator can therefore probe thermalising (scrambling) quantum behaviour  in 2D with the background curvature of a black hole. By analysing level statistics, we demonstrate that adding the background curvature strengthens the ergodicity expected in the flat lattice. 
 Together with the realisation of Hawking radiation, being able to implement such chaotic behaviour could enable us to realise the Hayden-Preskill quantum teleportation protocol responsible for the information retrieval from evaporating black holes \cite{hayden2007black, langlett2022rainbow, agarwal2023long}.
 
 Finally, combined with quantum gas microscopes, optical lattices naturally provide access to multi-particle correlations, thus offering an ideal system to monitor the behaviour of quantum information during the evaporation of black holes encoded in our simulator by changing the tunnelling coefficients in a time-dependent fashion.

\acknowledgements

We wish to thank Matthew Horner, F. Nur \"{U}nal, Mehedi Hasan and Dan E. Browne for useful discussions. A.B. acknowledges funding from the EPSRC Centre for Doctoral Training in Delivering Quantum Technologies at UCL, Grant No. EP/S021582/. G.N. acknowledges funding from the Royal Society Te Apārangi and the Cambridge Trust. U.S. acknowledges funding from the EPSRC Programme Grant DesOEQ (Grant No. EP/P009565/1) and the EPSRC Hub in Quantum Computing and Simulation EP/T001062/1. J.K.P. acknowledges support by EPSRC, Grant No. EP/R020612/1.

\bibliography{references.bib}

\appendix

\end{document}


\title{Supplementary Material\\Probing quantum properties of black holes with a Floquet-driven optical lattice simulator} 

\author{Asmae Benhemou}
\thanks{These authors contributed equally.}

\author{Georgia Nixon}
\thanks{These authors contributed equally.}

\author{Aydin Deger}

\author{Ulrich Schneider}

\author{Jiannis K. Pachos}

\maketitle

\section{Relating the tight-binding model to the Dirac Equation} Here we present a brief derivation showing how the dynamics of non-interacting atoms moving in an optical lattice can simulate the dynamics of a relativistic particle described by the Dirac equation. We start with the Heisenberg equation of motion for the fermionic annihilation operator $\hat{c}_j$ at a specific site $j$, i.e.
\begin{equation}
i \partial_t \hat{c}_j = [\hat{c}_j,\mathcal{H}],
\end{equation}
where fermionic anti-commutation relations $\{\hat{c}_j,\hat{c}_{j'}^\dagger\}=\delta_{j,j'}$ apply. The tight-binding Hamiltonian describing atoms moving in an optical lattice is given as
\begin{equation}
    \mathcal{H} = -\sum_{j} \left [ \kappa_j (\hat{c}_j^{\dagger}\hat{c}^{}_{j+1} + \hat{c}_{j+1}^{\dagger}\hat{c}^{}_j)  \right ],
\end{equation}
where we set $J_0=c=\hbar=1$ (see main text where $J_0$ is defined as the global energy scale).
It can be shown that $[\hat{c}_{j},\mathcal{H}]=-\kappa_j \hat{c}_{j+1} - \kappa_{j-1} \hat{c}_{j-1}$.
After substituting $\hat{c}_j=i^j \phi_j$ where the fermionic operator is transformed to the fermionic field $\phi$, we arrive at the following equation of motion for the (discrete) field $\phi_j$:
\begin{equation}
\partial_t \phi_j = - \kappa_j \phi_{(j+1)} + \kappa_{j-1} \phi_{(j-1)}
\label{eq:lateq}
\end{equation}
We next discuss the solution of the (1+1)D Dirac equation in the massless limit
\begin{equation}
i\gamma^a e^\mu_a (\partial_\mu +\omega_\mu)\psi = 0,
\end{equation}
which takes a similar form, allowing us to identify the relationship between the tunneling elements $\kappa_j$ of the lattice model  and 
a curved background with the metric $ds^2 = f(x)dt^2 - \frac{1}{f(x)}dx^2$. 
To be concrete, we consider a black hole with a horizon at $x=x_h$ and transform the Schwarzschild metric to Eddington-Finkelstein coordinates \cite{eddington1924comparison, finkelstein1958past}. In this metric the time coordinate remains constant along the radial null geodesics, resulting in non-singular metric components. The metric can be expressed as $ds^2 = f(x)dt^2 - 2dtdx$. Next, we select vierbeins as $e^0_t=[-f(x)-1]/2$, $e^1_t=[-f(x)+1]/2$, while the remaining vierbein components are equal to 1 and gamma matrices as $\gamma^0=\sigma^z$, $\gamma^1=i\sigma^y$ \cite{yang2020simulating}. The vierbeins are linking the curved metric $g_{\mu \nu} =\eta_{ab} e^a_\mu e^b_\nu$ to the flat metric $\eta_{ab}=\text{diag}(1,-1)$, while $\omega_\mu$ represents the spin connection. We select the following spinor $\psi=(\phi,-\phi)/\sqrt{2}$, where the two components are interdependent. Upon incorporating all these elements into the Dirac equation, we arrive at the formula \cite{yang2020simulating}
\begin{equation}
\partial_t \phi =-\frac{1}{4} \big\{ \partial_x \left[f(x) \phi \right] + f(x) \partial_x \phi \big\}.
    \label{eq:eq5}
\end{equation}
We now discretise Eq.~\eqref{eq:eq5} by replacing the continuous coordinate $x$ with a discretised version $jd$, where $j$ is an integer and $d$ denotes  the discretisation length. Together with the central difference formula for products of functions
\begin{equation}
\partial_x (f(x) \phi) = \frac{f_{j+1}\phi_{j+1}-f_{j-1}\phi_{j-1}}{2d},
\end{equation}
we obtain the following form
\begin{equation}
\partial_t \phi_j = - \left[\frac{f_{j+1}+f_{j}}{8d}\right] \phi_{(j+1)} + \left[\frac{f_{j-1}+f_{j}}{8d}\right] \phi_{(j-1)}.
\label{eq:direq}
\end{equation}
Now, by comparing Eq.~\eqref{eq:lateq} with Eq.~\eqref{eq:direq}, we obtain the relation from the main text

\begin{equation}
\kappa_j=\frac{f_{j+1}+f_j}{8d}.
\end{equation}

\section{Floquet Engineering}

The full time-dependent Hamiltonian describing a Floquet-driven 2D lattice system is given by
\begin{equation}
    H(t) = -J_0 \kappa \sum_{ \langle i,j \rangle} (\hat{c}_i^{\dagger} \hat{c}_{j} + \hat{c}_{j}^{\dagger} \hat{c}_i ) + \cos (\omega t) \sum_j A_j \hat{c}_j^{\dagger} \hat{c}_{j}.
    \label{Eq:TimeDependent1DHamiltonian_app}
\end{equation}
The equivalent of Eq.~\eqref{Eq:TimeDependent1DHamiltonian_app} in 1D is shown in Eq.~(4) in the main text. The stroboscopic Hamiltonian $H_S^{t_0}$ generates the evolution over one driving cycle from $t_0$ to $t_0 + T$. The choice of $t_0$ is a gauge choice, named the ``Floquet gauge". There exist a family of stroboscopic Hamiltonians $H_S^{t_0}$ for each $t_0 \in \{0, T \}$ and they are all related by a unitary transformation.

We can numerically calculate the stroboscopic Hamiltonian by simulating the evolution of time-independent basis elements under the Schr\"{o}dinger equation using the time-dependent Hamiltonian Eq.~\eqref{Eq:TimeDependent1DHamiltonian_app}. After one period, we diagonalise the time evolution operator $U(t_0 + T, t_0)|\psi_n(t_0)\rangle=u_n |\psi_n(t_0)\rangle$. The quasienergy $\{ \epsilon_n \}$ is given by the natural logarithm $\epsilon_n = \frac{i}{T} \log (u_n )$, defining the stroboscopic Hamiltonian via $H_S^{t_0} = \sum_n \epsilon_n |\psi_n(t_0)\rangle \langle\psi_n (t_0)|$. 

We can change basis and work with a representation of the stroboscopic Hamiltonian which is expressly $t_0$-independent. This Floquet-gauge invariant Hamiltonian is labelled the ``effective Hamiltonian," $H_{\mathrm{eff}}$ \cite{Eckardt-Anisimovas-HighfrequencyApproximationPeriodically-2015}. To explore the analytical behaviour of the effective Hamiltonian and utilise the high-frequency expansion \cite{Bukov-Polkovnikov-UniversalHighFrequencyBehavior-2015, Eckardt-Anisimovas-HighfrequencyApproximationPeriodically-2015}, it is convenient to make a gauge transformation from Eq.~\eqref{Eq:TimeDependent1DHamiltonian_app} into the ``lattice frame" given by
\begin{equation}
    \tilde{H}(t) = -J_0 \kappa\sum_{\langle i,j \rangle} e^{i\frac{A_i - A_j}{\hbar \omega} \sin (\omega t)} \hat{c}_i^{\dagger} \hat{c}_j,
    \label{Eq:HLatticeFrame}
\end{equation}
where the time dependence has been transferred to time-dependent Peierls phases \cite{Eckardt-ColloquiumAtomicQuantum-2017}.  

There exists an infinite series expansion, which describes the effective Hamiltonian involving terms of increasing order of the inverse frequency 
\begin{equation}
H_{\mathrm{eff}} = \sum_{n=0}^{\infty}  \frac{1}{\omega^n} H_{\mathrm{eff}}^{(n)}.
\label{Eq:Eq11_app}
\end{equation}
For sufficiently large driving frequencies, the effective Hamiltonian can be well approximated by the first two terms \cite{Eckardt-Anisimovas-HighfrequencyApproximationPeriodically-2015, Bukov-Polkovnikov-UniversalHighFrequencyBehavior-2015}
\begin{equation}
  H_{\mathrm{eff}}^{(0)} = \tilde{H}_0,  \quad \quad \quad 
     H_{\mathrm{eff}}^{(1)} = \frac{1}{\hbar } \sum_{l=1}^{\infty} \frac{1}{l} [\tilde{H}_l, \tilde{H}_{-l}] 
\end{equation}
where $\tilde{H}_l = \frac{1}{T} \int_0^T e^{-i l \omega t} \tilde{H}(t)$ are Fourier harmonics of the time-dependent Hamiltonian in the lattice frame given by Eq.~\eqref{Eq:HLatticeFrame}.
They are given by
$\tilde{H}_m = -J_0 \kappa \sum_{\langle i,j \rangle} \hat{c}_i^{\dagger} \hat{c}_j \mathcal{J}_m \left( \frac{A_i - A_j}{\omega} \right)$
where $\mathcal{J}_m$ is the Bessel function of the $m^{\mathrm{th}}$ kind.
The first term ($n=0$) of the high-frequency expansion Eq.~\eqref{Eq:Eq11_app} is therefore the first Fourier harmonic which corresponds to a Hamiltonian 
\begin{equation}
    H=-J_0\sum_{\langle i , j \rangle }  \kappa_{ij} (\hat{c}_i^{\dagger}\hat{c}^{}_{j} + \hat{c}_{j}^{\dagger}\hat{c}^{}_i) 
    \label{Eq:2DTightBinding}
\end{equation}
with tunnelling elements given by 
\begin{equation}
\kappa_{ij} = \kappa \mathcal{J}_0 \left( \frac{|A_i - A_{j}|}{\hbar \omega} \right),
\label{Eq:KappaFromDrive_app}
\end{equation}
see also the main text. We utilise this approximation to generate a sequence of local driving amplitudes that generate the desired tunnelling profile encoding the black hole geometry. The dimensionless constant $\kappa$ relates the chosen energy scale $J_0$ to the bare tunnelling in the un-driven optical lattice, i.e.\ the first term of Eq.~\eqref{Eq:TimeDependent1DHamiltonian_app}. We make an explicit distinction between these energy scales to enable a wider variety of Hawking temperatures in our simulator. The Bessel function in Eq.~\ref{Eq:KappaFromDrive_app} is limited to values between $1$ and $\sim-0.4$; the constant $\kappa$ acts as essential leverage to achieve larger effective tunnelling elements and therefore larger simulated $\alpha$ and encoded Hawking temperatures [see Eq.~(3) in the main text]. For instance, the analysis on error resilience in the main text simulates $\alpha$ values between $1$ and $3$ which would require $\kappa \gg 1$ for large lattices. 

Since $[\tilde{H}_m, \tilde{H}_{-m}]=0$, the second-order term of the high-frequency expansion vanishes
and the leading-order corrections to the effective Hamiltonian scale as $1/\omega^2$ as discussed in the main text.

Finally, we note that the complex phase of the tunnelling elements in a one-dimensional lattice constitutes a gauge freedom. Hence, a system with a purely real and positive ``V-shaped'' tunnelling profile i.e. the absolute value of the linear profile given in Eq.~3 in the main text would deliver the same dynamics given in Fig.~3 (main text) and provide an equally valid simulation of Hawking radiation. In contrast, tunnelling phases cannot be gauged away in 2D.

\section{Simulating a finite 2D system}

\begin{figure}
    \centering
    \includegraphics{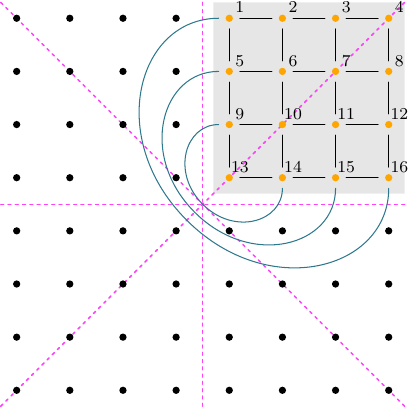}
    \caption{An $8\times 8$ lattice with $90\degree$ rotational symmetry can be simulated using $16$ sites. The black and yellow dots show the full $8\times 8 $ lattice, only the yellow dots in the shaded region are needed for the $16$ site simulation. Black lines indicate tunnelling between neighbouring sites common to both the full system and the $16$ site simulation. To the simulation, we add three additional tunnelling terms indicated by blue curves, connecting sites $1 \rightarrow 16$, $5 \rightarrow 15$, $9 \rightarrow 14$. }
    \label{fig:cone_boundary}
\end{figure}
To calculate the spectral, many-body properties of our 2D black hole simulator introduced in the main text, we simulate an $8 \times 8$ lattice. We utilise the $C_4$ ($90\degree$ rotational) symmetry of our simulator to simplify the numerical calculation. The $C_4$  symmetry allows us to use {\it cone boundary conditions} and quarter the number of sites needed. For example, Fig.~\ref{fig:cone_boundary} shows an $8 \times 8$ lattice which can be simulated using only the $16$ enumerated sites in the top right-hand corner of the lattice. This can be done by including three additional tunnelling terms as indicated by the blue curved links.

Our simulated black hole has $D_4$ symmetry: this includes $C_4$  rotational symmetry as well as mirror symmetry about each fuschia line in Fig.~\ref{fig:cone_boundary}. To remove the mirror symmetries, we add a randomly generated onsite energy term on site $3$ in our simulation in the main text.

\bibliography{references.bib}